\documentclass[pra,12pt]{revtex4-1}

\usepackage{graphicx}
\usepackage{amssymb} 
\usepackage{amsmath}    
\usepackage{latexsym}   
\usepackage{bm}
\usepackage{times}
\usepackage{centernot}
\usepackage{float}
\usepackage{color}
\usepackage{mathtools} 

\begin{document}
\title{Agency and the physics of numbers}
\author{John M. Myers}
\email{myers@seas.harvard.edu}
 \affiliation{Harvard School of Engineering and Applied 
Sciences, Cambridge, MA 02138, USA.}
\author{F. Hadi Madjid}%
 \email{gailmadjid@comcast.net}
 \affiliation{82 Powers Road, Concord, MA 01742\\ \today}

\begin{abstract} 
Analogous to G\"odel's incompleteness theorems is a theorem in physics to
the effect that the set of explanations of given evidence is uncountably
infinite.  An implication of this theorem is that contact between theory
and experiment depends on activity beyond computation and
measurement---physical activity of some agent making a guess. Standing on
the need for guesswork, we develop a representation of a symbol-handling
agent that both computes and, on occasion, generates a guess in
interaction with an oracle.  We show: (1) how physics depends on such an
agent to bridge a logical gap between theory and experiment; (2) how to
represent the capacity of agents to communicate numerals and other symbols,
and (3) how that communication is a foundation on which to develop both
theory and implementation of spacetime and related competing schemes for
the management of motion.
\end{abstract}

\maketitle
\section{Introduction}

Schr\"odinger, in his 1954 book ``Nature and the Greeks,'' laments \cite{NatGrk}:
\begin{quote}
\ldots  science in its attempt to describe and understand Nature
simplifies this very difficult problem.  The scientist subconsciously,
almost inadvertently, simplifies his problem of understanding Nature
by disregarding or cutting out of the picture to be constructed,
himself, his own personality, the subject of cognizance.
Inadvertently the thinker steps back into the role of an external
observer.  This facilitates the task very much.  But it leaves gaps,
enormous lacunae, leads to paradoxes and antimonies whenever, unaware
of this initial renunciation, one tries to find oneself in the picture
or to put oneself, one's own thinking and sensing mind, back into the
picture.  This momentous step \ldots has consequences.  So, in
brief, we do not belong to this material world that science constructs
for us.  We are not in it, we are outside.  We are only spectators. 
\end{quote}
A start toward recognizing the role in physics of the physicist can be found
in G\"odel's incompleteness theorems that show that mathematical logic in
uncloseably open to additional axioms.  G\"odel exploited the writing of
mathematics as strings of symbols, as did Turing in his definition of
computability.  More recently, we recognized that both evidence and its
explanations, as expressed in quantum theory, are written in strings of
symbols, and from this recognition we proved another incompleteness theorem,
to do with explanations in terms of wave functions and operators that fit
evidence expressed as parameter-dependent probabilities:
\begin{quote}
  {\bf Theorem:} The set of inequivalent explanations that exactly fit given
  probabilities is uncountably infinite \cite{18inc}.
\end{quote}
 Logically, there can be no unique explanation.  Of course, physicists
 collaborate and compete with one another to develop a dominant explanation.
 By virtue of the theorem however, the assertion of particular explanations
 reaches beyond logic; the assertion is a conjecture, an hypothesis, in
 short, a {\bf guess}, even if a community coalesces around it.  And the guess is
 forever subject to refutation, so that physics involves and endless open
 cycle in which physicists act as agents that make guesses \cite{huxQuote}.
 Here, not as an opinion but stemming from a proof within quantum theory, is
 a role in physics for the physicist.
 \newpage

\begin{figure}[h!]
  \hspace*{1 in}
    \includegraphics[height=3.45 in]{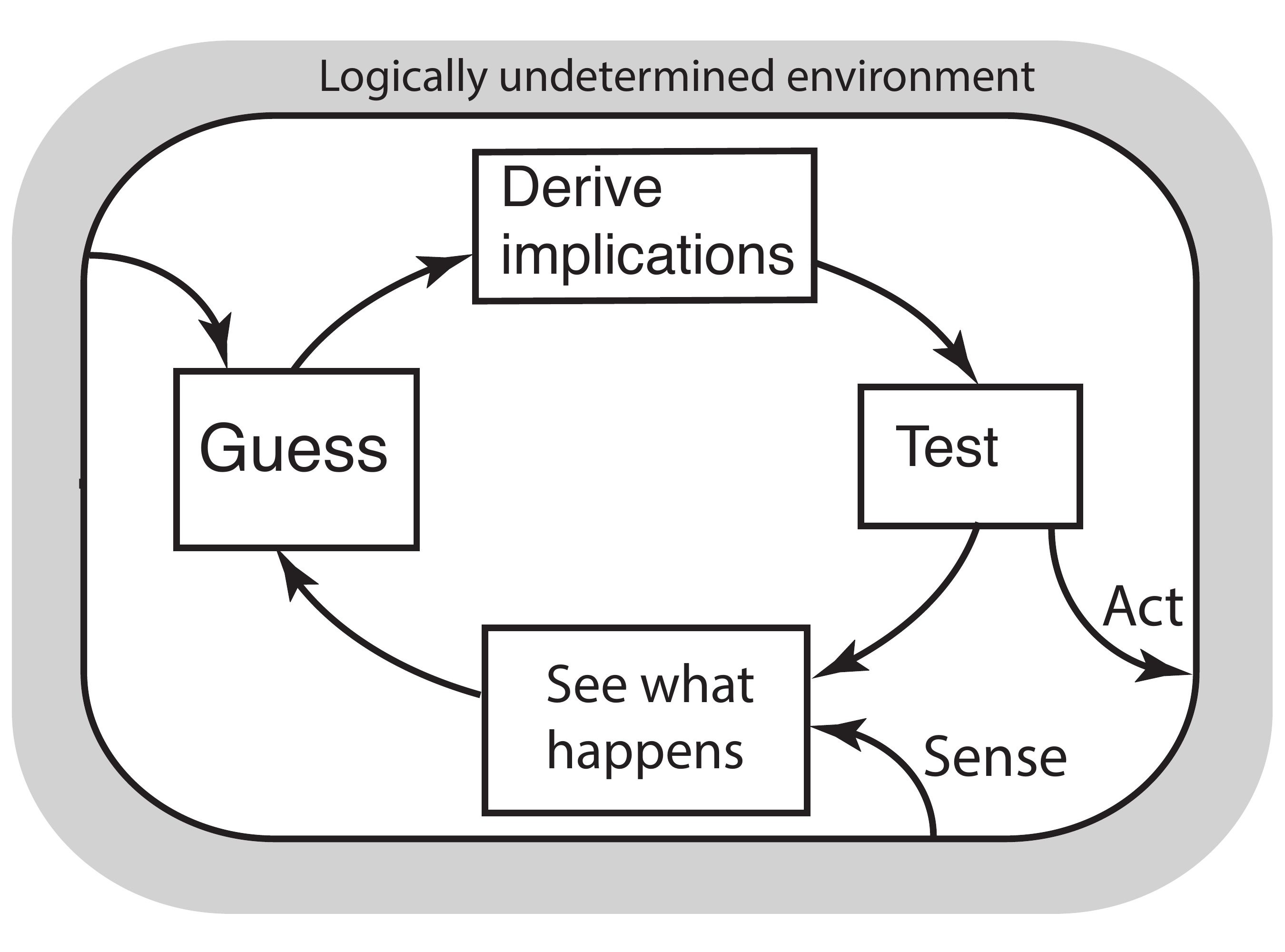}
    \caption{Agent's open cycle of guessing and testing}
  \end{figure}

Facing up to symbols and the agents that handle them in physical
investigations opens a whole new arena of physics, ready for exploration.
Agency in physics has been discussed from several points of view
\cite{97stapp,caves,notwithstand,briegel1104,06laskey,prakash2}.  We focus on
agents that perceive and use symbols.  But how, in the context of physics,
can one to speak of symbols?  Physics has sometimes been viewed as a quest
for ``fundamental elements'', mainly particles and fields, from which ``all
physical descriptions'' could be constructed; however, this quest has been
challenged, and in its place one can now enjoy more latitude, thanks in part
to notions of ``emergence'' made prominent in explaining phenomena of
condensed matter \cite{00pines}.  Availing ourselves of the latitude to take
symbols as elements of description from computer science and engineering,
here we advance notions of symbols and the agents that handle them, with
implications for theories of spacetime and their implementations.

\section{Agents perceive symbols}
In order to introduce into physics the notions of symbols and the agents that
perceive them, we need first to distinguish perception from measurement.
Here is a homely experiment on perception. I sit at a table and pick up a
magazine with pictures that I have not already looked at.  I shut my eyes and
turn the cover to expose the first page, still with my eyes shut. Then I
blink, very briefly opening my eyes and quickly shutting them again.  I have
an fleeting image of seeing a picture of a tiger.  Then I repeat the process.
Two or three glances later, the image stabilizes, not into a tiger by into a
woman sitting on a couch.  This little story illustrates the claim in Pattern
Theory of a two-way flow between brain and eye: the brain proposes forms that
the eye confirms or rejects \cite{mumford}.  In some circumstances, I can
make definite, stable perceptions, while outside these circumstances my
perceptions are unstable.  In circumstances in which I achieve a stable
perception, that perception might be wrong---as when confronted by an
illusion, as in the Ames demonstrations \cite{ames}, but my perception is not
uncertain in the way a measurement is, with its systematic and statistical
uncertainties.

In theoretical work at the blackboard, physicists perceive the
symbols---numerals, letters, plus and minus signs, etc.---in which they
write and read both evidence and its explanations.  But what about
experiments at the workbench in which they measure something?  To measure
something, one has to first perceive what is to be measured.  To measure
the voltage on a wire at a moment, a person must perceive the the
voltmeter, the wire to be measured, and the moment at which to make the
measurement, in distinction to other devices, other wires, and other
moments.

In particular, what about measuring the physical transmission of symbols?
Suppose Person A perceives the words in an email sent by Person B.  A third
Person C can measure the physical transmission of the email from A to B.
Person C may measure voltage levels at moments when these convey binary
symbols (bits) that code a higher level symbol (letter of the alphabet or
word), along with timing relations among various voltages.  Person C's
measurements produce numbers accompanied by uncertainties that contrast with
Person B's perception of letters on the computer screen.  The investigative
Person C who measures must also perceive.  For example, Person C has to
perceive both what is to be measured and the devices used to measure it.

To say that one {\it perceives} a symbol is to say that one registers
something, distinct from something else.  When perception matters, the symbol
perceived matters to ones next action, distinct from something else that
would contribute to a different action.  We will discuss circumstances
necessary to cases in which, apart from occasional failures, the perception
of symbols is unambiguous; for example, one perceives a dog as distinct from
a cat. While the cat you pet is more than a symbol, to recognize the cat as a
cat is to recognize `cat' as a symbol.

Two interrelated questions: what sort of entity has the capability to
perceive, and what sort of entity is it that is perceived?  Not just
people, but other living organisms make distinctions, e.g. between self and
non-self, and thus also perceive.  To perceive is to register a sensing in
some category and not in others.  Without an entity's capacity to make such
distinctions, the entity cannot be said to perceive.  We call such an
entity an {\it symbol-handling agent}, whether it is a living organism of a
machine.  For the second question---what sort of entity is perceived---we
propose that whatever is perceived is reasonably called a {\it symbol}.
Perception assigns a sensing to a category, without measuring it.  Such
categories are elements of thought, having much in common with mathematical
entities that are not measured but are defined in relation to other such
entities.  Words are symbols.  For perception as we mean it in this report,
humans perceive by labeling with words.  Wordless organisms that perceive
must have other ways of labeling, that is, of assigning symbols.

{\it Symbols} and {\it symbol-handling agents} are terms of description
available at widely varying levels of detail.  I may see myself as handling
symbols, and I may inquire into evidence of symbol handling on the part of
mantra within my cells.  A word as a symbol can be coded in letters
of an alphabet which in turn can be coded in bits.

What  capabilities, besides the perception of symbols, are to be ascribed to an `agent'?  I see a core capability of agency as a feature of the human condition: I
respond to surprises.  When the surprise comes to me, my sense is that it
comes in part from the unknowable but also in part from me, because I've been
working on something, and the surprise from the unknowable and the surprise
from me combine to come up with a guess.  We think of an agent as
taking steps, one after another, and as equipped with a memory.  Each ``next
step'' of an agent is influenced both by the contents of its memory and by an
inflow of symbols from an environment that includes other agents, and also by
a logically undetermined ``oracle'' external to the agent \cite{or}.  Guesses
emerge from an agent interacting with an oracle, the workings of which we
refrain from trying to penetrate.

\section{Steering rhythms of perception}
People delegate a variety of symbol-handling tasks to mechanized
agents, such as the computer at which I now write.  Although stripped of emotional capacity, dead, computer networks are important to understanding
symbol-handling agents in two ways.  First, human handling of symbols leaves
records in computer networks that serve as evidence.  Second, computer
networks exhibit rhythms of symbol handling important to agency in general.
Computer networks pervade the modern technological world, and not only in the
form of the Internet.  When I bring my car to the repair shop, the first
thing checked is often the computer network that mediates between my eyes,
hands, feet, and the mechanisms for steering, braking, and communication.
Records of mechanized agents visible in computer based records are a
heretofore overlooked topic for physical investigation.  The computers that
work as networked agents have much to teach us about the perception of
digital symbols, to do with rhythms of perception that require continual
maintenance, guided by responses to analog deviations that are logically
undetermined.

Analog machinery is idiosyncratic: two people using a slide rule get
slightly different answers. The huge strength of digital, as opposed analog,
computation comes from the capacity of a digital computer to make exact
calculations by means of inexact machinery.  Insofar as a person using the
computer is concerned, variations in the behavior of transistors leave a
computer's traffic in 0's and 1's unaffected.  This is because the basic unit
of memory, the flip-flop, works like a hinge that can be recognized
unambiguously as turned one way or the other, regardless of imperfections in
its manufacture and operation.  Or think of a chess game in a player
recognizes the placement of a pawn on a square of the board in spite of small
deviations of the pawn from the center of the square.  But, as we shall see,
recognizing symbols unambiguously is possible only within a system of analog
adjustment. The hand guided by the eye places the pawn more or less in the
middle of a square of a chess board---a performance in which noticing
variations is necessary to the adjustment of motion.

Analogous steering in the handling of digital symbols shows up with
special clarity in the physics of computer networks, such as the networks
used by human agents to mediate communications that generate evidence and
convey explanations.

\section{Mechanized symbol handlers}
In order to distinguish between what can be computed and what must come
from beyond computation (as guesses from interaction with an oracle), we
imagine the extreme case of a symbol-handling agent that, while open to
guessing, possesses maximal computational capacity. Unconcerned with
practical limits on computing imposed by limits on memory or by limits on
the rate at which an agent computes, we formulate a mechanical agent having
the computational capability of a Turing machine.  The Turing machine,
however, requires modification to offer a place for interaction with an
oracle and for communication with other such machines.

In his 1936 paper, Turing briefly introduced an alternative machine called a
{\em choice machine}, contrasted with the usual Turing machine that Turing
called an a-machine:
\begin{quote}
  If at each stage the motion of a machine \ldots is completely determined by
  the [memory] configuration, we shall call the machine an ``automatic
  machine'' (or a-machine). For some purposes we might use machines (choice
  machines or c-machines) whose motion is only partially determined by the
  configuration \dots. When such a machine reaches one of these ambiguous
  configurations, it cannot go on until some arbitrary choice has been made
  by an external operator.  This would be the case if we were using machines
  to deal with axiomatic systems. \cite{turing}.
\end{quote}
We picture a mechanical symbol-handling agent as implementing a c-machine
modified to take part in a communications network.  We call the c-machine
modified in this way, with some additional modifications to be described
shortly, a {\it Choice Machine}.  We posit that on occasion an ``oracle''
writes a symbol onto the scanned square of the Turing tape of the agent's
Choice Machine {\em privately}, in the sense that the symbol remains
unknown to other agents unless and until the Choice Machine that receives
the chosen symbol reports it to others \cite{aop16T}.  There is no limit to
the number of symbols that the oracle can write \cite{18inc}.

\section{Choice Machines paired by communications channels}\label{sec:4}
What affects one agent as a symbol need not be a symbol to
another agent, but there are special situations in which a pair of agents can
communicate by sending symbols back and forth from one agent to the other.
In this report we attend to the pairing of one agent with another agent by
means of a two-way communications channel over which symbols that mean
something to one agent arrive at another agent that coherently responds.

Characterizations of flow of symbols, including their necessary
synchronization, were discussed earlier \cite{aop14,qip15,aop16}.  Here we
discuss the mathematical forms---one might say data structures---that
underpin concepts of space, time, and spacetime as used by physicists, and
that also underpin arrangements of devices that implement these concepts.
We claim that these mathematical forms are important to thinking about how
organisms, human and non-human, manage motion.

\subsection{Symbol handling without metric}
To develop the theory of symbol handling to be presented below, we avoid
topics critical to the engineering of physical networks. Thus we give no
attention to how symbols might propagate from agent to agent, e.g. by
classical or quantum fields defined with respect to some coordinate system
with a metric tensor.  Rather, our approach is non-metric; one might say
topological.  We seek a theory of agents that maintain records of
transmissions and of receptions of symbols from one to another in a situation
that need have no coordinate system available.  Then there can be no central
clock, no assumption of any spacetime manifold and hence no metric tensor.
Indeed we must, so to speak, wipe these off the blackboard in order to get at
the element of novelty in attending to neworks of symbol-handling agents.
Eschewing any reference to {\it how} symbols are transmitted, propagated or
received, we postulate that symbol-handling agents register moments of
transmission and reception of symbols, each agent using its own count of
moments as a local clock.

\subsection{Forms of records used by agents to manage their motion }

To think about agents managing their motions, we explore cases in which  the Choice Machine expressing a symbol-handling capability is
augmented by a second tape that we call a {\it clock tape}.  Like the Turing
tape, the clock tape is marked into squares, each of which can hold a single
symbol, and at any moment a Choice Machine scans a single square of the clock
tape, but the motion of the clock tape never reverses: at each move the
scanned square of the clock tape shifts one place to the right.  Consider
symbols arriving one after another at an agent $A$ from an agent $B$.  Agent
$A$ can copy each symbol in the moment it arrives onto the scanned square of
$A$'s clock tape.  In this way $A$ maps the dynamic, temporal order of
arrival of the symbols to a static, spatial order along the clock tape.

Suppose that at some but not all of its moments, an agent $A$ receives a
symbol from $B$ and writes the received symbol on $A$'s clock tape.  If the
symbols arrive only occasionally, $A$ records these arriving symbols
sparsely, so that successive symbols occupy not successive squares of $A$'s
clock tape but on squares separated by stretches of $A$'s tape, stretches on
which $A$ may have written symbols for its own use or symbols arriving from
agents other than $B$.  Consider a stretch of $A$'s clock tape after such a
performance.  The ratio of the number of symbols recorded from $B$ to the
number of squares on $A$'s clock tape gives the average frequency ratio of
symbol arrivals from $B$ to steps of $A$.  Such frequency ratios are the
recorded evidence of the speed ratio of $A$'s receptions from $B$ relative
the speed of $A$'s stepping along its clock tape.  These ratios are the basic
form of evidence of motion.

  \begin{figure}[h!]
    \hspace*{1 in}
  \includegraphics[height=2.5 in]{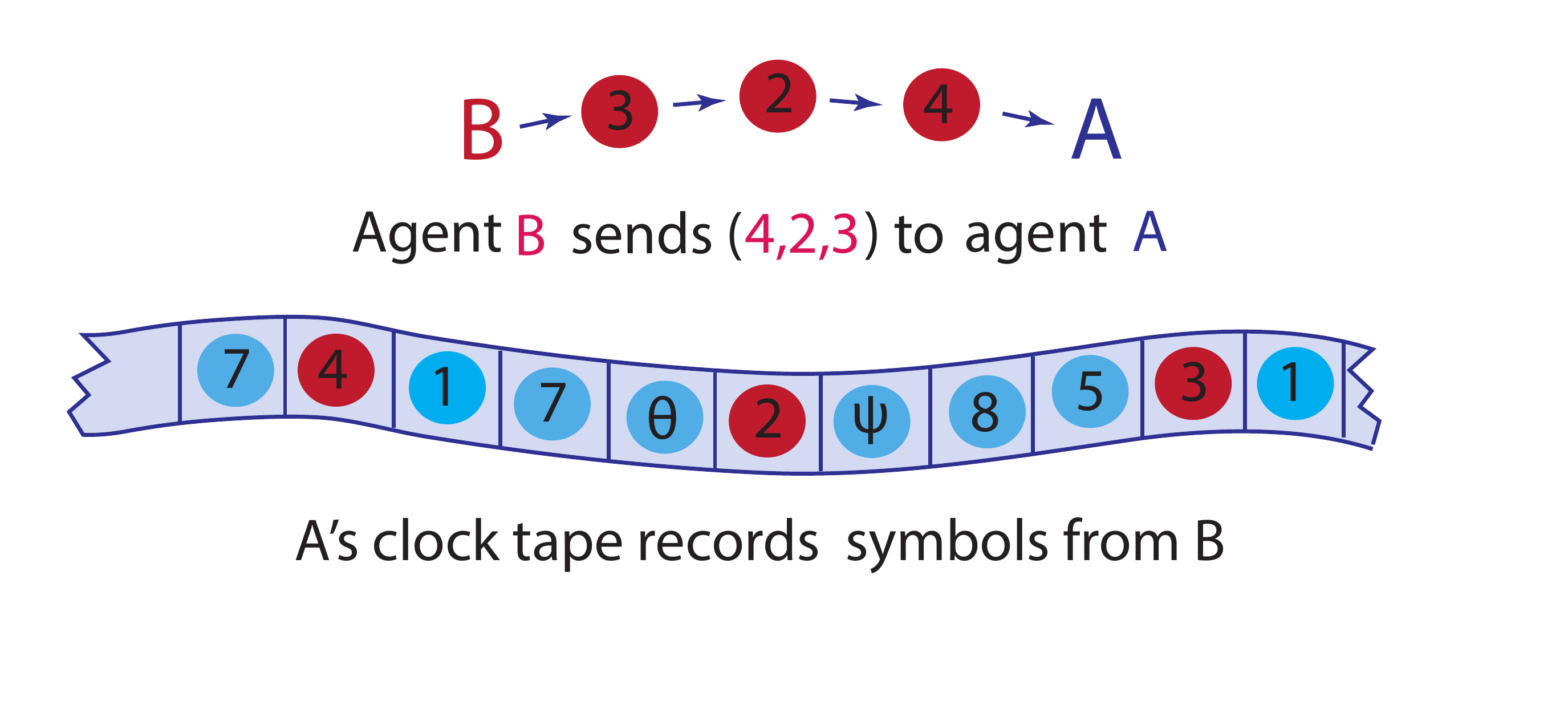}
  \caption{Clock-tape record of ``frequency''\label{fig:2}
  }
\end{figure}
  \noindent The example in Fig.~\ref{fig:2}
 shows a relative frequency
\[\frac{\text{number of }
  B\text{-symbols received}}{\text{number of }
  A\text{-moves}
}\approx \frac{1}{4}\]

Now we turn from frequency to {\it distance}. The basic form for distance in
terms of records on a tape is in terms of an {\it echo count}.  An echo count
is the number of squares from a symbol sent to the return of an echo.  That
is, agent $A$, at the moment of scanning square $n$ of its clock tape,
transmits a symbol to $B$, $B$ echoes the symbol back to $A$, and the echoed
symbol arrives as $A$ scans square $n+k$; then $k$ is an {\it echo count}
from $A$ to $B$ and back.

The notion of {\it echo count} is a basis for distance.  As discussed in more
detail in \cite{aop14,qip15}, in special cases echo counts it can be
aligned with distances as defined by Einstein in special relativity, but echo
count gives a property of records that is more basic than relativistic
distance, in that the concepts needs no assumption of any spacetime metric.

In discussing a lone agent, one might want to ask ``what belongs to the
agent?'', for example a record of odds that the agent assigns to a bet could
be said to ``belong to the agent.''  But when agents communicate, some
records are shared, so that ownership can be shared; communicating agents
differ drastically from lone agents.  We take as fundamental the capacity of
an agent, on occasion, to pair with another agent in communicating symbols.
Then the question ``what belongs to an agent?'' becomes embedded in a context
of asking ``what strings of symbols flow from one agent to another?''.
Strings of symbols bind the agents without belonging totally to one or the
other agent.  It is a potentially interesting discipline to explore this and
related questions, not just in engineered systems but also in organisms,
human and non-human, to explore situations in which one can ask: ``which
record, what agents have access to the record, where and when (along their
own tapes) do they have this access?''

\subsection{Evolution of forms of records}
The analysis of the form of records of frequency and distance has the
following implication. Given that agents, by definition, can be seen as in
interaction with oracles that are logically undetermined, and given that
records of symbols in the memories of agents are the basic form to frequency
and distance, we arrive at the insight that agents managing speed and
distance operate not as a single system of a logic of symbols landing on and
departing from the squares of tapes, but like extended arithmetics, they form
an uncountable set, involving the evolution through the logically
undetermined entrance of symbols, e.g. from oracles.

We are interested in various possible paths of the evolution of ways in which
agents manage motion, especially in cases in which agents leave measurable
tracks in the form of their clock tapes.  For example, human agents develop
evolving national and international time standards.  One such thread of
evolution consists in the concepts and the implementation of these concepts
in International Atomic Time (TAI).  Recall that TAI rests on its base
reference frequency of the electromagnetic radiation resonant with the
transition between hyperfine energy levels of cesium 133 \cite{aop14}.  This
reference frequency is employed, either directly or indirectly, in almost all
physical investigation.  But Cesium (or its contemplated replacement by an
optical frequency) is by no means the only type of reference imaginable or
potentially fruitful.  Other such threads of evolution of symbolic
communication among agents are candidates for exploration, and some are in
use.  For example, the cutting-edge optical clocks now operating in National
Metrology Laboratories have an instability of about 1/100 of the cesium
clocks that define the second.  Therefore it makes no sense to compare two
such optical clocks by comparing each to the second: one compares them
directly.  There is an interesting potential for alternative schemes of
measuring motion based on symbol exchange in the Laser Interferometer
Gravitational-Wave Observatory (LIGO) experiment \cite{ligo2016}.  Still
other schemes are candidates for explorations of the management of motion by
biological organisms, both human and non-human.  ``The frog starves to death
in the presence of dead flies, not because he is fastidious, but because he
does not see them'' \cite{lettvin}.  Some recent crashes of self-driving cars
have been traced to ``not seeing still objects.'' What alternative to
spacetime is it in which the frog lives?

\section{Discussion}
While we needed to start, as in \cite{18inc}, with a ``lone agent'',  a lone agent is like ``one hand clapping'': Symbol-handling agents can't function without being paired with other such agents via back-and-forth communication of symbols.  As discussed in \cite{aop14}, conflicts in synchronization place severe limitations the potential for a given agent to communicate with other agents, and the potential for conflicts generally limits the number channels that can operate concurrently to pair one agent with another.  Agent $A$ may have to step out of communication with $B$ if it is to enter communication with $C$.

As we see it, the notion of objectivity has to do with expectations of agreement among records made by different agents, rather than a notion of behavior devoid of agent participation.

The clock tape along which an agent steps is the core image of ``local
time,'' without encumbering it with the notion, which for 40 years we have
found baffling, of any ``global time'' \cite{aop14}.

Guesses by an agent, here thought of as arising in the interaction of an
agent with an oracle, influence not only how an agent resolves certain
choices but also the agent's organization; that is, the agent's capacity to
perceive a situation that calls for a choice.

\section*{Acknowledgment}
We thank Kenneth Augustyn, Paul Benioff, and Louis Kauffman for instructive  comments.

\end{document}